\documentclass[journal,10pt]{IEEEtran}
\usepackage{subfigure}
\PassOptionsToPackage{subfigure}{tocloft}
\usepackage{CJK}
\usepackage{algorithm}
\usepackage{algorithmic}
\usepackage{amsmath}
\usepackage{amssymb}
\usepackage{psfrag}
\usepackage{graphicx}
\usepackage{epstopdf}
\usepackage{multirow}
\usepackage{stfloats}
\usepackage{cite}
\usepackage{color}
\usepackage[normalem]{ulem}
\usepackage{arydshln}
\usepackage{enumerate}
\usepackage{bbm}
\usepackage{bm}

\newtheorem{mypro}{\bf{Proposition}}

\title{Model-Free  Optimization for Reconfigurable Intelligent Surface with Statistical CSI}
\author{
Huayan Guo,   Ying-Chang Liang, \IEEEmembership{Fellow, IEEE}, Sa Xiao
\thanks{This work was supported by the National Natural Science Foundation of China under Grant U1801261, 61631005, and 61571100.  (\textit{Corresponding author: Sa Xiao.})}
\thanks{The authors are with the National Key Laboratory of Science and Technology on Communications, University of Electronic Science and Technology of China (UESTC), Chengdu 611731, China, also with the Center for Intelligent Networking and Communications (CINC), University of Electronic Science and Technology of China (UESTC), Chengdu 611731, China (e-mail: guohuayan@pku.edu.cn; liangyc@ieee.org;xiaosajordan23@163.com).}
}

\begin{document}

\maketitle

\begin{abstract}
In this paper, a single user multiple input single output downlink wireless communication system is investigated, in which multiple reconfigurable intelligent surfaces (RISs) are deployed to improve the  propagation condition.
Our objective is to optimize the phase shift matrices of all the RISs by exploiting the statistical channel state information (CSI).
In particular, two model-free algorithms are proposed, which are applicable for any channel  statistical assumptions.
Numerical results show that the proposed algorithms   significantly outperform the random phase shift scheme, especially when the channel randomness is low.
\end{abstract}


\section{Introduction}
\emph{Reconfigurable intelligent surface} (RIS) is a passive radio technique, which can intentionally tune the electromagnetic behavior of the wireless environment with meta-materials \cite{Liaskos2018magzineIRS,survey2019,WuQQmagzine,Liang2019JCIN,Debbah2019IRSMISO}.
Recently, RIS has been considered as a promising technique to  enhance the quality-of-service of users
with low power consumption and deployment cost. 
Specifically, most existing works investigate the joint optimization of the transmitter at the \emph{access point} (AP) and the phase matrix of the RIS, while assuming perfect \emph{channel state information} (CSI) is available. In these works, the key design challenge  is the non-convex unit-modulus constraint on the reflection coefficient of the RIS element.
By using the similar mathematical  tools  for analog precoding problems in traditional massive multiple input multiple output systems, joint optimization algorithms have been developed for transmit power minimization problem \cite{zhangrui2018GcomIRS,zhangruiIRS}, energy efficiency problems \cite{Yuen2018ZF,YuenChauIRS}, weighted sum-rate maximization problems \cite{WSRGuo,WSRmulticell}, and secrecy rate maximization problems \cite{zhangruiwcl,Chenjie2019access} in RIS-aided systems.

However, the perfect CSI is generally unavailable in practice, since the passive RIS  has no capability  to sense the channel.
For this reason, it is more reasonable to optimize the RIS with statistical CSI.
In \cite{Jinshi2019TVTsurfaceAverage}, the statistical CSI RIS setup is firstly investigated for the single user \emph{multiple input multiple output} (MISO)  system, 
and the phase matrix is optimized to maximize the average received signal power at the user.
In \cite{Dennah2019IRS}, a multi-user system is investigated, where the AP-RIS channel is assumed known, and
the max-min fairness algorithm is designed based on the random matrix theory.
Nevertheless, the algorithms in \cite{Jinshi2019TVTsurfaceAverage} and \cite{Dennah2019IRS} are designed for specific  channel models. When the channel model changes, new algorithms are still required.

In this paper, a single-user MISO system with multiple RISs is investigated. We aim at designing model-free phase matrix optimization methods under the statistical CSI setup, which can work under any channel  statistical assumptions.
Based on the \emph{stochastic successive convex approximation} (SSCA) technique in \cite{Palomar2016ssca} and \cite{LiuATSP2018onlineSSCA}, a stationary-solution-achieved algorithm is designed to maximize the average achievable rate.
To facilitate implementation, an online implementable algorithm is further designed based on the \emph{stochastic majorization-minimization} (SMM) method \cite{ZQLuo2016SSUM} to optimize the average receive \emph{signal to noise ratio} (SNR) instead of the average achievable rate.
Simulation results verify that the online implementable algorithm can achieve a comparable performance in comparison with the stationary-solution-achieved algorithm. In addition, both proposed algorithms significantly outperform the random RIS phase scheme.

\begin{figure}
[!t]
\centering
\includegraphics[width=.5\columnwidth]{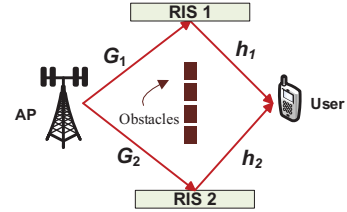}
\caption{An RIS-aided  multiuser MISO communication system.}
\label{IRS_system}
\end{figure}

\section{System Model}\label{system model}
This paper investigates a RIS-aided MISO communication system as shown in {\figurename~\ref{IRS_system}}.
The AP is equipped with $M$ antennas, and serves a single-antenna user.
The direct transmission link between the AP and the user is blocked by the obstacles.
In order to improve the propagation condition, $K$ RISs are deployed to provide high-quality virtual links from the AP to the user.
Suppose each RIS has $N$ reflection elements.
The baseband equivalent channels from the AP to the $k$-th RIS, and from the $k$-th RIS to the user at time slot $i$ are denoted by ${\bf G}_{k,i} \in {\mathbb C}^{N\times M}$ and ${\bf h}_{k,i} \in {\mathbb C}^{N \times 1}$, respectively.
Then, the received signal at the user is expressed by
\begin{equation}\label{equ:downlink_signal_1}
{y_i}=
\left(\sum_{k=1}^K {\bf h}_{k,i}^{\rm H} {\bf \Theta}_k {\bf G}_{k,i} \right) {\bf w}_i s_i +u_i
,
\end{equation}
where ${\bf \Theta}_k={\rm diag}(\vartheta_{1,k}, \cdots, \vartheta_{n,k}, \cdots, \vartheta_{N,k})$ is the diagonal phase-shift matrix of the $k$-th RIS with $\vartheta_{n,k}= e^{\jmath \varphi_{n,k}}$, $u_i \sim {\cal{CN}}(0,\sigma_0^2) $ denotes the \emph{additive white Gaussian noise} (AWGN) at the receiver, $s$ is the information-bearing signal at AP with unit power, and ${\bf w}_i \in {\mathbb C}^{M\times 1}$ is the corresponding transmit beamforming vector at the AP.
We further define
\begin{equation*}
\begin{aligned}
{{\bm \theta}}=
\begin{bmatrix}
{\rm diag}({\bf \Theta}_{1}^{\rm H}) \\
\vdots\\
{\rm diag}({\bf \Theta}_{K}^{\rm H})
\end{bmatrix}
, \;
{{\bf H}_i}=
\begin{bmatrix}
{\rm diag}({\bf h}_{1,i}^{\rm H}){\bf G}_{1,i} \\
\vdots\\
{\rm diag}({\bf h}_{K,i}^{\rm H}){\bf G}_{K,i}
\end{bmatrix}
.
\end{aligned}
\end{equation*}
Then the received signal ${y_i}$ is equivalently represented by
\begin{equation}\label{equ:downlink_signal_2}
{y_i}= {\bm \theta}^{\rm H} {\bf H}_i{\bf w}_i s_i +u_i
.
\end{equation}

According to \cite{TSE}, the optimal transmit beamforming can be expressed as the following function of ${\bf H}_i^{\rm H} {\bm \theta}$
\begin{equation}\label{equ:MRT0}
{\bf w}_i=\sqrt{P_{\rm T}} \frac{{\bm \theta}^{\rm H} {\bf H}_i  } {\|{\bm \theta}^{\rm H} {\bf H}_i\|}
,
\end{equation}
where $P_{\rm T}$ is the transmit power constraint.
Therefore, the instantaneous achievable rate at time slot $i$ is
\begin{equation}\label{equ:Ri}
\begin{aligned}[b]
R_i({\bm \theta})&=\log \left(1+\frac{P_{\rm T} }
{\sigma_0^2} {\left\|{\bm \theta}^{\rm H} {\bf H}_i\right\|^2}\right)
.
\end{aligned}
\end{equation}

It is known that the passive RIS has no capability of channel estimation.
In addition, the dimensions of ${\bf h}_{k,i}$ and ${\bf G}_{k,i}$ makes channel estimation very challenging, especially when $N$ is large.
As a result, it is difficult to obtain the perfect knowledge of these channel coefficients in most practical cases.
In this paper, we propose to design ${\bm \theta}$ only based on the statistical information of ${\bf h}_{k,i}$ and ${\bf G}_{k,i}$, and the optimization problem for the statistical CSI setup can be formulated as follows:
\begin{align*}
{\mathcal{P}}{\text{(A)}} \quad \max_{ \forall {\bm \theta}} \quad & {\mathbb E}_{i} \left[R_i({\bm \theta}) \right] \notag \\
{\bf s.t.} \quad
& |\theta_{n}| =1, \quad \forall n=1,\cdots,KN,
\end{align*}
where ${\mathbb E}_{i} \left[R_i({\bm \theta}) \right]$ is the expectation of $R_i({\bm \theta})$ over all time slot.

It should be noted that the above problem is independent of the statistical models of channels. Thus it is more general and practical than model-based studies in \cite{Jinshi2019TVTsurfaceAverage} and \cite{Dennah2019IRS}.

\section{Stochastic Successive Convex Approximation for Average Rate Maximization}
Generally speaking, there are two main challenges to solve ${\mathcal{P}}{\text{(A)}}$.
Firstly, similar to the perfect CSI-based studies in \cite{zhangrui2018GcomIRS,zhangruiIRS,Yuen2018ZF,YuenChauIRS,WSRGuo,WSRmulticell,zhangruiwcl,Chenjie2019access}, the  non-convex unit-modulus constraint is intractable.
Secondly, the objective function ${\mathbb E}_{i} \left[R_i({\bm \theta}) \right]$ is non-convex, and usually cannot be denoted by closed-form expression.

To deal with these challenges, we design a stationary-solution-achieved algorithm for ${\mathcal{P}}{\text{(A)}}$ in this section.
At first, we replace $\bm \theta$  by $\bm \varphi$ to make the unit-modulus constraint a convex form, where ${\theta}_n=e^{\jmath { \varphi}_n}$ and ${\bm \varphi}=[\varphi_1,\cdots,\varphi_{NK}]^{\rm T}$.
Then,
the optimization for the non-convex objective function is addressed by the SSCA technique \cite{Palomar2016ssca,LiuATSP2018onlineSSCA}.

\subsection{Optimization Variable Substitution}
Replacing $\bm \theta$ by $\bm \varphi$, the rate expression becomes
\begin{equation}\label{equ:Ri_phi}
\begin{aligned}[b]
{\bar R}_i({\bm \varphi})&=\log \left(1+\frac{P_{\rm T} }
{\sigma_0^2} {\left\|{\left(e^{\jmath {\bm \varphi}}\right)}^{\rm H} {\bf H}_i\right\|^2}\right)
.
\end{aligned}
\end{equation}
Thus ${\mathcal{P}}{\text{(A)}}$ is translated to
\begin{align*}
{\mathcal{P}}{\text{(A1)}} \quad \min_{ \forall {\bm \varphi}} \quad & f({\bm \varphi})=-{\mathbb E}_{i} \left[{\bar R}_i({\bm \varphi}) \right] \notag \\
{\bf s.t.} \quad
& \varphi_n \in {\mathbb R}, \quad \forall n=1,\cdots,KN.
\end{align*}
Since the objective function is continuously differentiable with Lipschitz continuous gradient, ${\mathcal{P}}{\text{(A1)}}$ is the standard stochastic social optimization problem satisfying the applicable assumptions in \cite{Palomar2016ssca}.

\subsection{SSCA for ${\mathcal{P}}{\text{(A1)}}$}
${\mathcal{P}}{\text{(A1)}}$ is still difficult to solve, since the objective function is non-convex with no closed-form expression. Fortunately, these issues can be dealt with SSCA technique.
The key idea of the SSCA technique is first to approximate the gradient of $f({\bm \varphi})$ with a carefully-designed incremental sample estimate, and then linearize the non-convex part of $f({\bm \varphi})$ with the obtained gradient to render a convex surrogate objective function.
Specifically, when the random realization ${\bf H}_i$ at time slot $i$ is obtained, ${\bm \varphi}$ is updated by the following three-step method: 
\begin{itemize}
\item Firstly, we approximate the gradient of $f({\bm \varphi})$ with a incremental simple estimate. Define ${\bf f}^{(i)}$ and ${\bm \varphi}^{(i)}$ to be the approximated gradient and the obtained ${\bm \varphi}$ at time slot $i$, respectively. Then we have
\begin{equation}\label{equ:grad_hat_f}
{\bf f}^{(i)}=(1-\rho^{(i)}) {\bf f}^{(i-1)}-\rho^{(i)} \nabla {R}_i({\bm \varphi}^{(i-1)}),
\end{equation}
where $\rho^{(i)}=i^{-\beta}$, $0.5\le\beta\le1$, and $\nabla { R_i}({\bm \varphi})$ is the gradient of $R_i({\bm \varphi})$. According to the chain rule, we have
\begin{equation}
\nabla { R}_i({\bm \varphi})={\rm Re} \left\{ -\jmath {\bm \theta}^\ast \circ  \nabla R_i({\bm \theta}) \right\},
\end{equation}
where  $\circ$ denotes the Hadamard product, ${\bm \theta}^\ast$ is the  conjugate of ${\bm \theta}$, and define ${\bf A}_i=\frac{P_{\rm T} } {\sigma_0^2} {\bf H}(i) {\bf H}(i)^{\rm H}$, we have
\begin{equation}
\nabla R_i({\bm \theta})=  \frac{2 {\bf A}_i {\bm \theta}}{1+{\bm \theta}^{\rm H}{\bf A}_i{\bm \theta}}.
\end{equation}

\item Secondly, we find the incremental element of ${\bm \varphi}$ at time slot $i$. Based on the obtained ${\bf f}^{(i)}$, $f({\bm \varphi})$ is approximated by the following surrogate function:
\begin{equation}\notag
\begin{aligned}[b]
{\hat f}_i({\bm \varphi},{\bm \varphi}^{(i-1)})
=\langle{\bm \varphi}-{\bm \varphi}^{(i-1)},{\bf f}^{(i)}\rangle+\frac{\tau}{2} \|{\bm \varphi}-{\bm \varphi}^{(i-1)}\|^2,
\end{aligned}
\end{equation}
where $\tau>0$, and $\langle{\bf x},{\bf y}\rangle$ is the inner product of vectors $\bf x$ and $\bf y$.
Note that ${\hat f}_i({\bm \varphi},{\bm \varphi}^{(i-1)})$ is a convex function. Thus the incremental element can be obtained by solving the following convex optimization problem
\begin{align*}
{\hat{\bm \varphi}}  =\arg \; \min_{ \forall {\bm \varphi}} \; & {\hat f}_i({\bm \varphi},{\bm \varphi}^{(i-1)}),
\end{align*}
and its closed-form expression is
\begin{equation}\label{equ:hat_varphi}
{\hat{\bm \varphi}}={\bm \varphi}^{(i-1)}-\frac{{\bf f}^{(i)}}{\tau}.
\end{equation}
\item Finally, ${\bm \varphi}$ is updated by
\begin{equation}\label{equ:hat_var}
{\bm \varphi}^{(i)}=(1-\gamma^{(i)})  {\bm \varphi}^{(i-1)}+\gamma^{(i)} {\hat{\bm \varphi}},
\end{equation}
where $\gamma^{(i)}=i^{-\alpha}$, and we need $\beta<\alpha\leq 1$ to guarantee convergence \cite{Palomar2016ssca}.
\end{itemize}

We summarize the proposed algorithm in  {\emph{Algorithm} \ref{alg:P3}}.
Despite the concise updating rules, the hyperparameter $\tau$ should be fine-tuned to realize a fast converge speed as well as a good performance, which requires extra off-line signal processing cost when  the statistical CSI changes \cite{Palomar2016ssca}.
In the next section, we will proposed an online implementable method,which can obtain a sub-optimal solution.

\begin{algorithm}[!ht]
\caption{ SSCA for Average Rate Maximization.}
\label{alg:P3}
\begin{algorithmic}[1]
\STATE {Initialize ${\bm \varphi}^{(0)}$ and $\epsilon$, and set ${\bf f}^{(0)}={\bf 0}$ and $i=0$.\\
{\bf Repeat}}
\STATE Obtain new channel realization ${\bf H}_i$;
\STATE Update ${\bf f}^{(i)}$ according to \eqref{equ:grad_hat_f};
\STATE  Solve ${\hat{\bm \varphi}}$ by \eqref{equ:hat_varphi};
\STATE Let ${\bm \varphi}^{(i)}=(1-\gamma^{(i)})  {\bm \varphi}^{(i-1)}+\gamma^{(i)} {\hat{\bm \varphi}}$;
\STATE $i=i+1$;
\\
{\bf Until} $\left|{\hat f}_i({\bm \varphi}^{(i)},{\bm \varphi}^{(i-1)})-{\hat f}_i({\bm \varphi}^{(i-1)},{\bm \varphi}^{(i-1)})\right|<\epsilon$.
\end{algorithmic}
\end{algorithm}

\section{Stochastic Majorization-Minimization for Average SNR Maximization}
One can see that, the main design challenge to resolve ${\mathcal{P}}{\text{(A)}}$ is the logarithm function in $R_i({\bm \theta})$, which makes the objective function neither convex nor concave.
In this section, a heuristic but implementation friendly algorithm is proposed, which approximates the average rate by the average received SNR.\footnote{It has been shown in \cite{Jinshi2019TVTsurfaceAverage} that, to maximize the average received SNR is a tight approximation of the original average achievable rate maximization problem.}
Specifically, we have following approximated problem:
\begin{align*}
{\mathcal{P}}{\text{(B)}} \quad \min_{ {\bm \theta} } \quad &
g({\bm \theta})\triangleq
{\mathbb E}_{i} \left[ g_i({\bm \theta}) \right] \notag \\
{\bf s.t.} \quad
& |\theta_{n}| =1, \quad \forall n=1,\cdots,KN,
\end{align*}
where $g_i({\bm \theta})={\bm \theta}^{\rm H} {\bf B}_i {\bm \theta}$, and ${\bf B}_i=-{\bf H}(i) {\bf H}(i)^{\rm H}$.
Although both the constraint set and the objective function of ${\mathcal{P}}{\text{(B)}}$ are still non-convex, we will show that the constraint set can be relaxed to the convex one equivalently, and then iterative algorithm with simple online-implementable updating rule for ${\mathcal{P}}{\text{(B)}}$ can be designed based on the SMM method \cite{ZQLuo2016SSUM}.

\subsection{Unit-Modulus Convex Relaxation}
The objective function $g({\bm \theta})$ could be further written as
\begin{equation}
g({\bm \theta})={\bm \theta}^{\rm H} {\bar{\bf B}} {\bm \theta},
\end{equation}
where ${\bar{\bf B}}={\mathbb E}_{i} \left[ {\bf B}_i \right]$.
Then, based on following proposition, the non-convex phase constraint can be relaxed.
\begin{mypro}\label{app_convex}
The optimization problem ${\mathcal{P}}{\text{(B)}}$ is equivalent to the following problem with convex constraint set:
\begin{align*}
{\mathcal{P}}{\text{(B1)}}\quad \min_{{\bm \theta}} \quad &  g(\bm \theta) \notag \\
{\bf s.t.} \quad
& |\theta_{n}|^2 \leq 1, \quad \forall n=1,\cdots,KN.
\end{align*}
\end{mypro}

\begin{IEEEproof}
Since ${\bf B}_i =-{\bf H}(i) {\bf H}(i)^{\rm H}$ is negative semidefinite,
${\bar{\bf B}}$ is also negative semidefinite, and $g(\bm \theta)$ is concave.
Therefore, in ${\mathcal{P}}{\text{(B1)}}$, the optimal $\theta_{n}$ is chosen at the boundary of the constraint set $|\theta_{n}|^2 \leq 1$.
\end{IEEEproof}
Proposition \ref{app_convex} reveals the fact that,
every element on the RIS should adopt the highest reflection strength to maximize the average (or instantaneous) SNR.

\subsection{SMM for ${\mathcal{P}}{\text{(B1)}}$}
SMM is an iterative algorithm to deal with the non-convex stochastic optimization problem like ${\mathcal{P}}{\text{(B1)}}$, which is the combination of the \emph{sample average approximation} (SAA) method and the \emph{ majorization-minimization} (MM) method \cite{MM2017Palomar}.

According to the SAA method, in the $t$-th time slot, the realization ${\bf B}_t$ is coming, and the output ${\bm \theta}^{(t)}$ is updated by solving:
\begin{align*}
{\mathcal{P}}{\text{(B2)}}\quad {\bm \theta}^{(t)}&=\arg \;  \min_{{\bm \theta}} \;  {\tilde g}_t({\bm \theta})\triangleq \frac{1}{t} \sum_{i=1}^t g_i({\bm \theta}) \notag \\
{\bf s.t.} \quad
& |\theta_{n}|^2 \leq 1, \quad \forall n=1,\cdots,KN.
\end{align*}
Denote the SAA of ${\bf B}_i$ as follows:
\begin{equation}\label{hat_B}
{\tilde{\bf B}}_t= \frac{1}{t} \sum_{i=1}^t {\bf B}_i
,
\end{equation}
and  then ${\tilde g}_t({\bm \theta})$ becomes 
\begin{equation}
{\tilde g}_t({\bm \theta})={\bm \theta}^{\rm H} {\tilde{\bf B}}_t {\bm \theta}.
\end{equation}

However, since $g_i({\bm \theta})$ is non-convex, ${\tilde g}_t({\bm \theta})$ is non-convex as well, and ${\mathcal{P}}{\text{(B2)}}$ is challenge to solve.
Fortunately, a stationary-solution achieved method has been proposed in \cite{ZQLuo2016SSUM} for this kind of problem, which is the stochastic extension version of the conventional MM method.
In particular, stationary solution for ${\mathcal{P}}{\text{(B1)}}$ can be  obtained by solving the following approximation problem iteratively:
\begin{align*}
{\mathcal{P}}{\text{(B3)}}\quad {\bm \theta}^{(t)}&=\arg \; \min_{{\bm \theta}} \;  {\bar g}_t({\bm \theta})\triangleq  \frac{1}{t} \sum_{i=1}^t {\hat g}_i({\bm \theta},{\bm \theta}^{(i-1)}) \notag \\
{\bf s.t.} \quad
& |\theta_{n}|^2 \leq 1, \quad \forall n=1,\cdots,KN.
\end{align*}
where ${\hat g}_i({\bm \theta},{\bm \theta}^{(i-1)})$ is the surrogate function for $g_i({\bm \theta})$, which is uniformly strongly convex, and satisfies the MM constraint \cite{MM2017Palomar}:
\begin{subequations}
\begin{align}
&{\hat g}_{i}({\bm \theta}, {\bm \theta}) = {g}_{i}({\bm \theta}), \label{equ:sca_con1}\\
&{\hat g}_{i}({ {\bm \theta}}, {\bm \theta}^{(i-1)}) \geq {g}_{i}({ {\bm \theta}}). \label{equ:sca_con2}
\end{align}
\end{subequations}

In the next, we first design surrogate function which satisfies \eqref{equ:sca_con1} and \eqref{equ:sca_con2}. Then, updating rule is designed by solving ${\mathcal{P}}{\text{(B3)}}$.

\subsubsection{Surrogate Function}

Since ${\bf B}_i \preceq 0$, we have $\left({\bm \theta}-{{\bm \theta}^{(i-1)}}\right)^{\rm H} {\bf B}_i \left({\bm \theta}-{{\bm \theta}^{(i-1)}}\right)\leq 0
$. Based on that, one may design surrogate function as follows:
\begin{equation}\label{equ:surrogate}
\begin{aligned}[b]
{\hat g}_{i}({\bm \theta},{{\bm \theta}^{(i-1)}})
&= {{\bm \theta}^{(i-1)}}^{\rm H} {\bf B}_i {{\bm \theta}^{(i-1)}}
-2 {\rm Re} \left\{{{\bm \theta}^{(i-1)}}^{\rm H} {\bf B}_i {\bm \theta}\right\}\\
&\qquad\qquad+\tau \|{\bm \theta}-{{\bm \theta}^{(i-1)}}\|^2
.
\end{aligned}
\end{equation}
Note that, equations \eqref{equ:sca_con1} and \eqref{equ:sca_con2} hold even though $\tau=0$. Thus we may set $\tau$ as an arbitrary small number to make ${\hat g}_{i}({\bm \theta},{\bm \theta}^{(i-1)})$  uniformly strongly convex.

\subsubsection{Problem Decompose}
Define ${\bf d}_t$ as the  SAA of ${\bf B}_i {\bm \theta}^{(i-1)}$, which is updated recursively:
\begin{equation}\label{equ:d_t}
{\bf d}_t=\frac{1}{t} {\bf B}_t {\bm \theta}^{(t-1)}+(1-\frac{1}{t}) {\bf d}_{t-1}
.
\end{equation}
Then we have
\begin{equation}\label{equ:d_t2}
\frac{1}{t} \sum_{i=1}^t {\rm Re} \left\{{{\bm \theta}^{(i-1)}}^{\rm H} {\bf B}_i {\bm \theta}\right\}
={\rm Re} \left\{ {\bf d}_t^{\rm H} {\bm \theta}\right\}.
\end{equation}
Substituting \eqref{equ:d_t2} into  ${\bar g}_t({\bm \theta})$, and removing the constant terms, ${\mathcal{P}}{\text{(B3)}}$ is equivalently written by
\begin{align*}
{\mathcal{P}}{\text{(B3)}}\quad {\bm \theta}^{(t)}&=\arg \; \min_{{\bm \theta}} \;  {\ddot g}_t({\bm \theta}) \notag \\
{\bf s.t.} \quad
& |\theta_{n}|^2 \leq 1, \quad \forall n=1,\cdots,KN,
\end{align*}
where ${\ddot g}_t({\bm \theta})=
-2 {\rm Re} \left\{ {\bf d}_t^{\rm H} {\bm \theta}\right\}+\frac{1}{t}\sum_{i=1}^t \tau \|{\bm \theta}-{{\bm \theta}^{(t-1)}}\|^2$.
In addition, ${\ddot g}_t({\bm \theta})$  can be further written by
\begin{equation}\label{equ:obj_PB3_decouple}
{\ddot g}_t({\bm \theta})=
\sum_{n=1}^{NK} {\ddot g}_{n,t}({ \theta_n})
,
\end{equation}
where $ {\ddot g}_{n,t}({ \theta_n})$ is the function of ${ \theta_n}$:
\begin{equation}
 {\ddot g}_{n,t}({ \theta_n})
 = -2 {\rm Re} \left\{ {d}_{n,t}^{\rm H} {\theta_n}\right\}+\frac{1}{t}\sum_{i=1}^t \tau \|{\theta_n}-{\theta_n^{(t-1)}}\|^2
,
\end{equation}
and $d_{n,t}$ is the $n$-th element of ${\bf d}_t$.
Therefore, based on \eqref{equ:obj_PB3_decouple}, problem ${\mathcal{P}}{\text{(B3)}}$ can be decoupled into $KN$ subproblems with respect to each $\theta_{n}$:
\begin{align*}
{\mathcal{P}}{\text{(B4)}}\quad \theta_n^{(t)}&=\arg \; \min_{{ \theta_n}}  \;
 {\ddot g}_{n,t}({ \theta_n})
 \notag \\
{\bf s.t.} \quad
& |\theta_{n}|^2 \leq 1.
\end{align*}

\subsubsection{Updating Rule}
${\mathcal{P}}{\text{(B4)}}$ is the convex optimization problem, which can be solved by the \emph{Karush-Kuhn-Tucker} (KKT) conditions.
Define the Lagrangian as
\begin{equation}
{\cal L}(\theta_n,\lambda)={\ddot g}_{n,t}({ \theta_n})+\lambda (|\theta_{n}|^2-1)£¬
\end{equation}
where $\lambda\geq 0$.
Then according to the KKT conditions, the optimal $\theta_n$ and $\lambda$ should satisfy
\begin{subequations}
\begin{align}
&\nabla_{\theta_n} {\cal L}(\theta_n,\lambda)=0, \label{equ:kkt_con1}\\
& \lambda (|\theta_{n}|^2-1)=0. \label{equ:kkt_con3}
\end{align}
\end{subequations}

From \eqref{equ:kkt_con1}, the optimal solution of ${\mathcal{P}}{\text{(B4)}}$ given $\lambda$ is:
\begin{equation}\label{equ:theta_single_PSCI_lambda}
\theta_n^{(t)}=\frac{d_n+\tau {\ddot \theta}_n^{(t)}}{\tau+\lambda},
\end{equation}
where ${\ddot \theta}_n^{(t)}$ is the  SAA of $\theta_n^{(i-1)}$, which can also be learnt recursively:
\begin{equation}\label{equ:bar_theta_nt}
{\ddot \theta}_n^{(t)}=\frac{1}{t} \theta_n^{(t)}+(1-\frac{1}{t}) {\ddot \theta}_n^{(t-1)}
.
\end{equation}

The rest task is to determine $\lambda$ based on \eqref{equ:kkt_con3} and \eqref{equ:theta_single_PSCI_lambda}.
Since we choose $\tau\approx 0$, when $\lambda=0$, constraint $|\theta_{n}|^2 \leq 1$ does not hold.
Thus we should have  $|\theta_{n}|^2=1$.
Finally, the optimal solution of ${\mathcal{P}}{\text{(B4)}}$ is:
\begin{equation}\label{equ:theta_single_PSCI}
\theta_n^{(t)}=e^{-\jmath \angle (d_n+\tau {\ddot \theta}_n^{(t)})}.
\end{equation}

The above SMM based algorithm is summarized in {\emph{Algorithm} \ref{alg:P2}}.
Note that, although  {\emph{Algorithm} \ref{alg:P2}} also has the parameter $\tau$, the performance and the convergence speed is not sensitive to the value of $\tau$, since equations \eqref{equ:sca_con1} and \eqref{equ:sca_con2} always hold for any $\tau$.
Since no off-line fine-tuning is required, {\emph{Algorithm} \ref{alg:P2}} can be deployed online, meanwhile the statistical CSI is learnt from the incoming observations without any prior knowledge about the channel.

\begin{algorithm}[!ht]
\caption{ SMM for Average SNR Maximization.}
\label{alg:P2}
\begin{algorithmic}[1]
\STATE {Initialize ${\bm \theta}^{(0)}$, and set  $t=1$.\\
{\bf Repeat}}
\STATE Obtain new channel realization ${\bf B}_t$;
\STATE Update ${\tilde{\bf B}}$, ${\bf d}_t$, and ${\ddot \theta}_n^{(t)}$ for all $n$ based on \eqref{hat_B}, \eqref{equ:d_t}, and \eqref{equ:bar_theta_nt}, respectively;
\STATE Update ${\bm \theta}^{(t)}$ according to \eqref{equ:theta_single_PSCI}.
\STATE $t=t+1$;
\\
{\bf Until} $\left|{\tilde  g}({\bm \theta}^{(t)})-{\tilde  g}({\bm \theta}^{(t-1)})\right|<\epsilon$.
\end{algorithmic}
\end{algorithm}

\section{Numerical Results}\label{simulation}
In this section, numerical examples are provided to validate the effectiveness of the proposed algorithms.
Suppose that the AP and the RISs are deployed properly, following \cite{Jinshi2019TVTsurfaceAverage}, the channels between the AP and the RISs are modeled in Rician fading, i.e.,
\begin{equation*}
{\bf G}_k= \sqrt{\frac{\rho}{\rho+1}} {\bar{\bf G}}_k+ \sqrt{\frac{1}{\rho+1}} {\tilde{\bf G}}_k,
\end{equation*}
where ${\bar{\bf G}}_k$ and ${\tilde{\bf G}}_k$  are the LoS and NLoS components, respectively, $\rho=10$ is the Rician factor,
and the elements of ${\tilde{\bf G}}_k$ are i.i.d. standard complex Gaussian distributed.
Same as \cite{Jinshi2019TVTsurfaceAverage}, we assume the multiple antennas at the AP and the reflector elements at the RIS are arranged in the \emph{uniform linear array} (ULA) for simplicity. Then, the LoS components are expressed by the responses of the ULA:
\begin{equation*}
{\bar{\bf G}}_k={\bm \alpha}_N(\phi_{{\text AoA},k}) {\bm \alpha}_M(\phi_{{\text AoD},k})^{\rm H},
\end{equation*}
where ${\bm \alpha}_N(\phi)=[1,e^{\jmath \pi \sin \phi},\cdots,e^{\jmath \pi (N-1) \sin \phi}]$, $\phi_{{\text AoA},k}$ is the AoA at the $k$-th RIS, and $\phi_{{\text AoD},k}$ is the AoD at the AP to the $k$-th RIS.
We further assume that there are $L=5$ channel paths between the RIS and the user. Then, the channel vector ${\bf h}_k$ is expressed as
\begin{equation*}
{\bf h}_k= \sum_{l=1}^L \beta_{l,k} {\bm \alpha}_N(\phi_{l,k})
,
\end{equation*}
where $\beta_{l,k} \sim {\cal{CN}}(0,\sigma_{\beta_{l,k}}^2) $, and $\sigma_{\beta_{l,k}}^2$ is randomly generated from an exponential distribution and normalized by $\sum_{l=1}^L \sigma_{\beta_{l,k}}^2=1$.
We set the transmission bandwidth  as $200$ kHz and noise power spectral density as $-170$ dBm/Hz.
In addition, suppose the distances between AP and RISs, and the distances between RISs and the user are the same, i.e., $d=10$ m.
Then, the path loss is modeled by $38.46+20 \lg d$ dB.

In simulation, we first generate $100$ snapshots, in which all the angles $\phi_{{\text AoA},k}$, $\phi_{{\text AoD},k}$ and $\phi_{l,k}$ are are randomly chosen from $(0, 2 \pi]$, and $\sigma_{\beta_{l,k}}^2$ is randomly generated from the exponential distribution and then normalized. 
Then, in each  snapshot, we further generate $100$ independent realizations of small-scale fading ${\tilde{\bf G}}_k$ and $\beta_{l,k}$. 
We set $\epsilon=0.01$ as the stoping criterion for the proposed algorithms.

\begin{figure}
[!t]
\centering
\includegraphics[width=.92\columnwidth]{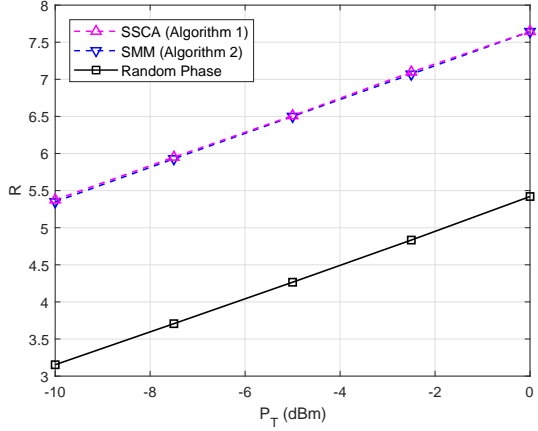}
\caption{R versus $P_{\rm T}$, when $M=4$, $K=2$, and $N=20$.}
\label{R_vs_PT}
\end{figure}

{\figurename~\ref{R_vs_PT}} illustrates the average rate of different schemes with respect to the transmit power $P_{\rm T}$.
One can see that, the proposed algorithms may achieve about $10$ dB gain compared with the  random RIS phase scheme.
In addition, {\emph{Algorithm} \ref{alg:P3}} and \ref{alg:P2} achieve almost the same performance.
Hence, optimizing the average received SNR is a good approximation to optimizing the average achievable rate, which is coincident with the conclusion in \cite{Jinshi2019TVTsurfaceAverage}.

\begin{figure}
[!t]
\centering
\includegraphics[width=.92\columnwidth]{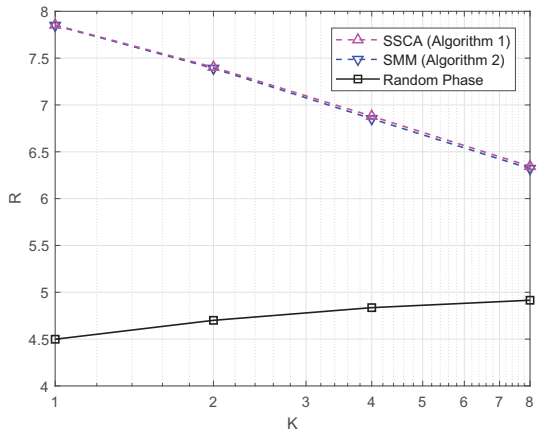}
\caption{R versus $K$, when $M=4$, $P_{\rm T}=-5$ dB, and $N K=64$.}
\label{R_vs_K}
\end{figure}

Then, in {\figurename~\ref{R_vs_K}}, the total RIS element number $NK$ is fixed to $64$, and the average achievable rate is plotted for different $K$.
Since each RIS has different $\phi_{{\text AoA},k}$, $\phi_{{\text AoD},k}$, and $\phi_{l,k}$, when $K$ increases, the performance of the random RIS phase scheme increases slightly due to the diversity gain.
However,  the performance of the statistical-CSI based schemes decreases drastically, since the number of the long-term variables required to be learnt increases proportionally to $K$.
As a result, in practice, we think it is better to serve one user with only one RIS.

\section{Conclusion}\label{conclusion}
In this paper, we investigate the multiple-RIS-aided downlink MISO system.
Two algorithms are developed to optimize the phase matrices of the RISs by exploiting the statistical CSI.
Both algorithms are applicable for any channel model assumptions.
Numerical results  verify that the proposed algorithms may sufficiently outperform the random phase scheme, especially when $K$ is small.

\bibliographystyle{IEEEtran}
\bibliography{IEEEabrv,mybib_draft2}

\end{document}